\documentstyle[namedreferences]{kluwer}
\input psfig.sty

\def\refe{\par\noindent\hangindent=1.5cm}
\newenvironment{squote}{\begin{quote}\small\sl}{\end{quote}}

\begin{opening}

\title{Cosmological Constant and the Final Anthropic Hypothesis}

\author{Milan M.\ \surname{\'Cirkovi\'c}}

\institute{Astronomical Observatory, Volgina 7, 11000 Beograd, SERBIA \\
Dept. of Physics \& Astronomy, SUNY at Stony Brook, \\
Stony Brook, NY 11794-3800, USA \\
email: {\tt mcirkovic@aob.aob.bg.ac.yu}}

\author{Nick \surname{Bostrom}}

\institute{Dept. of Philosophy, Logic and Scientific Method \\
London School of Economics, Houghton St., WC2A AE, London, UK \\
email: {\tt n.bostrom@lse.ac.uk}}

\date{}
\end{opening}
\begin{document}

\begin{abstract}
The influence of recent detections of a finite vacuum energy 
("cosmological constant") on our formulation of anthropic 
conjectures, particularly the so-called Final 
Anthropic Principle is investigated. It is shown that non-zero 
vacuum energy implies the onset of a quasi-exponential 
expansion of our causally connected 
domain ("the universe") at some point in the future, a stage similar 
to the inflationary expansion at 
the very beginning of time. The transition to this future inflationary phase of cosmological expansion will preclude indefinite survival of intelligent species in our domain, because of the rapid shrinking of particle horizons and subsequent depletion of energy 
necessary for information processes within the horizon of any observer. Therefore, to satisfy the Final Anthropic Hypothesis (reformulated to 
apply to the entire ensemble of universes), it is necessary to show that (i) chaotic inflation of Linde (or some similar model) provides a satisfactory description of reality, (ii) migration between 
causally connected domains within the multiverse is physically permitted, and (iii) the time interval left to the onset of the future inflationary phase is sufficient for development of the technology necessary for such inter-domain travel. These stringent requirements diminish the probability of the Final Anthropic Hypothesis being true.
\end{abstract}

\section{Introduction}
The term "anthropic principle" was first introduced in the famous exposition of Carter (1974). The Strong Anthropic Principle (henceforth SAP), as defined by Carter, states that:
\begin{squote}
...the Universe (and hence the fundamental parameters on which it depends) 
must be such as to admit the creation of observers within it at some stage.
\end{squote}
This principle expresses an observational selection effect: If we have a cosmological theory according to which there are multiple universes or "domains", then when thinking about the observational consequences of 
that theory we have to take into account only those domains in which observers exist. For example, even if the vast majority of domains are inhospitable to life, the theory would predict that we will be 
observing one of the exceptional domains where conditions appear to 
be fine-tuned for life to exist. The only speculative feature of SAP 
is thus that it is applicable only if there is a multiverse; beyond 
that it is simply a logical truism, albeit an important one (Leslie 1989).
By contrast, the so-called Final Anthropic Principle, introduced in the 
monograph by Barrow \& Tipler (1986), can make no such claim to methodological status. It states that:
\begin{squote}
Intelligent information-processing must come into existence in the universe, and, once it comes into existence, it will never die out.
\end{squote}
We wish to remove the teleological overtones from this statement and 
focus on the latter part of it, which clearly expresses a scientific hypothesis that is susceptible to refutation on observational grounds. 
To avoid confusion, one may use the expression "Final Anthropic 
Hypothesis" (henceforth FAH) to refer to the following conjecture:
{\it Once intelligent information processing comes into existence, 
it will never die out.}

It should be noted at this stage that in their explanation and 
discussion of the Final Anthropic Principle, Barrow and Tipler (1986)
do not refer to multiple universes in any form. This should be 
emphasized all the more strongly for the fact that, in other parts 
of their book, the possibility that the observed world is but one among 
many others (the general idea of an ensemble of universes, a "multiverse") {\it is\/} explored. In light of recent data, as we shall show in the next 
section, the existence of a multiverse appears 
to be the only chance of indefinite survival of intelligent life. It is therefore essential 
to understand the FAH in such a way as to make room for this possibility. 
Even with this clarification, however, the FAH as stated above is susceptible 
to various differing interpretations. In particular, we need to distinguish between the 
following two meanings:
\begin{enumerate}
\item
There is at least one intelligent race in the universe that will continue to exist indefinitely. For the sake of brevity, we shall call this interpretation {\it individualistic}.

\item
Any particular intelligent race might eventually die out, but intelligent life as a whole will exist indefinitely. This interpretation may be termed {\it holistic}.
\end{enumerate}
Statement 1. logically implies statement 2., but not vice versa. In a sense, one could 
make a comparison with the local character of conservation laws in classical physics. As 
discussed by Feynman in his popular exposition (Feynman 1965) electric charge 
could, in principle, be conserved in two ways: locally and non-locally. Charge could 
be moved within a box, or it could vanish at one point and be created ex nihilo at 
another point within a box. Physics allows only for the former way of charge 
conservation, but laws governing life and intelligence in the universe are much less 
known than laws of classical physics. Obviously, the local way of conservation is 
analogous to the individualistic interpretation of the FAH conjecture, 
while non-local appearance and disappearance of charges are similar to rises and falls 
of intelligent communities at various points in spacetime.

One reason why FAH is an interesting hypothesis is that we might be 
interested in knowing whether there is any theoretical possibility for the human species or its descendants to survive indefinitely, in the sense of performing an infinite number of computations along its world-line 
(Tipler 1994). For the settlement 
of this question, the individualistic reading is obviously more relevant than the 
holistic reading. From now on, we shall therefore assume the individualistic reading 
when referring to the FAH, unless otherwise stated. 
Our goal is thus to investigate whether recent progress in cosmology can 
actually support the conjecture that information processing is necessarily finite within our (or any!) domain, and what recourse 
is left for the proponents of the FAH (see also Tipler 1986). In this respect, it seems that we are in the middle 
of a major change of cosmological paradigm (not unexpected, however, as even the cursory look at relevant literature could show). Recent 
results of the surveys of the Type I supernovae at cosmological distances strongly indicate the presence of a large cosmological constant 
(Perlmutter et al.~1998; Reiss et al.~1998). If the total 
cosmological density parameter corresponds to the flat ($\Omega=1$) universe, the contribution due to vacuum energy density is
\begin{equation}
\label{jedan}
\Omega_\Lambda \approx 0.7. 
\end{equation}
This result indicates not only that the universe will expand indefinitely, but that it will 
expand in an (asymptotically) exponential manner, 
manner, similar to the early inflationary phase in its history. It 
should be mentioned that a non-zero cosmological constant has been considered 
desirable from a several points of views in recent years, because it would be capable 
of solving the cosmological age problem and because it would arise naturally from 
quantum field theory processes (e.g.~Klapdor 
\& Grotz 1986; Singh 1995; Martel, Shapiro \& Weinberg 1998). 
A universe with $\Omega \approx 1$ and $\Omega_\Lambda$ similar to the value in (\ref{jedan}) would allow the 
formation of galaxies (e.g.~Weinberg 1987; Efstathiou 1995), and it could last long 
enough for the life, including intelligent observers, to evolve, all in agreement with 
SAP. However, we shall show that a universe with a cosmological constant of this magnitude is one where intelligent life cannot survive indefinitely, thus apparently contradicting FAH.

\section{Future of the $\Lambda$-universe}
Universes with a cosmological constant $\Lambda$, a subclass 
of the Friedman-Robertson-Walker (FRW) models, are represented 
by solutions of the Einstein field equations in the generalized form
\begin{equation}
\label{polik}
R_{\mu \nu} - \frac{1}{2} g_{\mu \nu} R - \Lambda g_{\mu \nu} = - \frac{8\pi G}{c^4} T_{\mu \nu},
\end{equation} 
where $\Lambda$ is a positive scalar (for other notation see any of 
the standard General Relativity textbooks, e.g.~Weinberg [1972]; 
for history and phenomenology of the cosmological constant, 
see the detailed review of Carroll, Press \& Turner [1992], and 
references therein). The value of $\Lambda$ in eq.~(\ref{polik}) is 
related to the vacuum cosmological density fraction as
\begin{equation}
\label{bazar}
\Omega_\Lambda = \frac{c^2 \Lambda}{3H_0^2} = 2.8513 \times 10^{55} 
\, h^{-2} \Lambda,
\end{equation}
where $\Lambda$ is in units of cm$^{-2}$, and $H_0$ is the Hubble 
constant ($H_0 \equiv 100\: h$ km s$^{-1}$ Mpc$^{-1}$). 
The Friedmann equation for the matter-dominated case can be written as
\begin{figure} 
\psfig{file=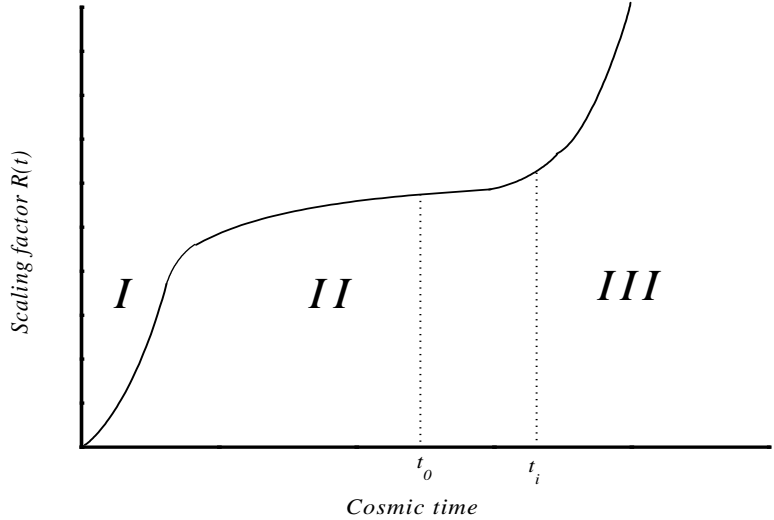,height=11cm,width=9cm}

\vspace{-2.5cm}

\caption{A sketch of evolution of a cosmological 
domain (conventional "universe"), with present epoch denoted with $t_0$, and major phases in the universal expansion labeled by Roman numerals (not drawn to scale). \label{auto2}}
\end{figure}
\begin{equation}
\label{fried1}
H(t) = \frac{\dot{a}(t)}{a(t)} = H_0 \left\{ \Omega_m \left[ \frac{a(t)}{a(t_0)} \right]^{-3} \!
+ \Omega_\Lambda - (\Omega_m + \Omega_\Lambda -1)\left[ \frac{a(t)}{a(t_0)} \right]^{-2} \right\}^{\frac{1}{2}}.
\end{equation}
In this formula, as well as in all others unless explicitly stated, cosmological matter density fraction $\Omega_m$ and vacuum energy density fraction $\Omega_\Lambda$ are evaluated at the present 
time. The quantity $K_0 = \Omega_m + \Omega_\Lambda -1$ is sometimes 
called normalized present-day curvature. For practical purposes, it is 
convenient to translate the eq.~(\ref{fried1}) into the "redshift 
language" as
\begin{equation}
\label{revi}
H (z) = H_0 [(1+z)^3 \Omega_m -(\Omega_m + \Omega_\Lambda -1)(1+z)^2
+ \Omega_\Lambda]^{\frac{1}{2}}.
\end{equation}
This relation governs the universal expansion and can be integrated into the past or future. Doing the latter shows that our domain
cannot remain matter-dominated at all times. Since the simplest dimensional analysis (as well as physical experience) shows that the vacuum energy density becomes progressively more important as the distances between fundamental observers increase, we expect the universe to become dominated by cosmological constant at some moment of its history. This epoch at 
which universe becomes vacuum-dominated is obtained from the equation (e.g.~Adams \& Laughlin 1997)
\begin{equation}
\label{al1}
t_i = t_0 + \tau_m \frac{\sinh^{-1} (1) - \sinh^{-1} (\sqrt{x})}{\sqrt{x}},
\end{equation}
(here $\sinh^{-1}$ denotes the inverse hyperbolic sine, not the reciprocal) where 
\begin{equation}
\label{dodat1}
x \equiv \frac{\Omega_\Lambda}{\Omega_m} = 
\frac{\Omega_\Lambda}{\Omega - \Omega_\Lambda} = \frac{\Omega_\Lambda}{1 -\Omega_\Lambda},
\end{equation}
the last equality, applying, of course, to the globally flat universe ($\Omega =1$). The current age of the universe is denoted by $t_0$ 
and the time constant $\tau_m$ is defined as
\begin{equation}
\label{al2}
\tau_m \equiv (6 \pi G \rho_m)^{-\frac{1}{2}} = \left( \frac{9}{4} H_0^2 \Omega_m \right)^{-\frac{1}{2}} = \frac{6.5327 \times 10^9 \; h^{-1}}{\sqrt{\Omega_m}} 
 \; {\rm yr}.
\end{equation}
Here, $\rho_m$ is the matter density of the universe (i.e.~matter plus - quite negligible for the purposes of the present discussion - radiation energy density). One should keep in mind that the current age of the universe for $\Omega_\Lambda$ given by the eq.~(\ref{jedan}) is (Perlmutter et al.~1998)
\begin{equation}
\label{cage}
t_0 = 14.9^{+1.4}_{-1.1} \times \frac{0.63}{h} \;{\rm Gyr}. 
\end{equation}
The history of our domain is schematically presented in the Figure \ref{auto2} (see also the complementary Fig.~1 in Kardashev 1997). Three major epochs are denoted by large Roman numerals. The epoch of primordial inflation (I) has been discussed in 
many - already classic - references (e.g.~Guth 1981; Linde 1983; Barrow 1988), and we shall not discuss its details here. The entire history of the universe, as we know it from observations (characterized mainly by the structure formation and evolution processes) is contained in the epoch of power-law expansion (II), described in classical cosmology textbooks (Weinberg 1972; Peebles 1993). We are currently inhabiting this expansion phase, our epoch being denoted by $t_0$; as can be seen from the eqs.~(\ref{jedan}) and (\ref{al1}) above, it is likely that $t_0$ is already located within the phase III, or at least close to the end of 
the phase II (see Fig. 2). 

\begin{figure} 
\psfig{file=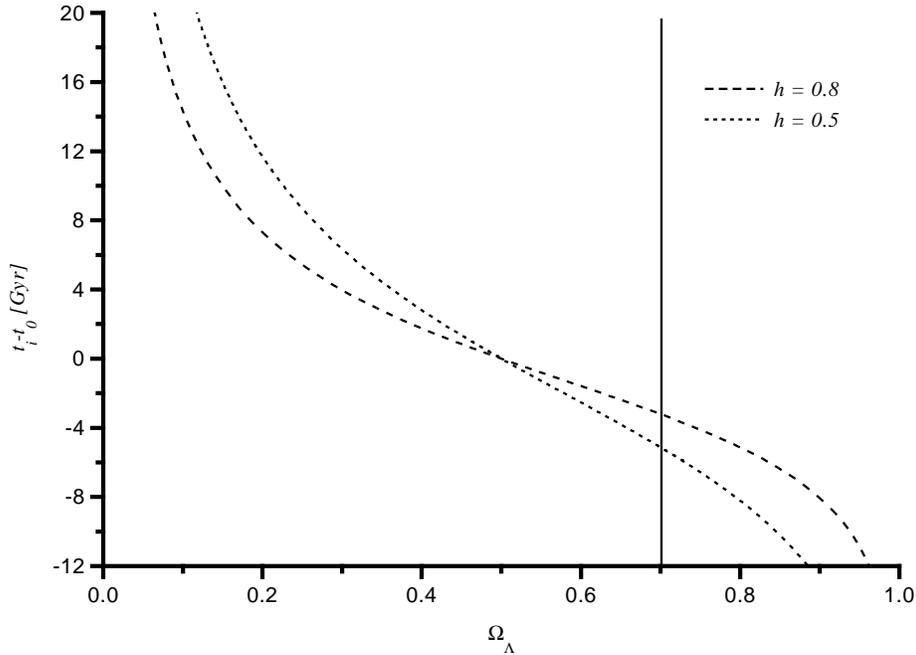,width=12cm,angle=270}
\caption{Duration of remaining time prior to the onset of future (quasi) exponentially expanding phase in Gyr as a function of cosmological 
constant contribution to the universal energy density in globally flat FRW universe. Vertical line represents the currently believed value of vacuum energy density. 
Note that 
that for $\Omega_\Lambda > 0.5$, the value of $t_i =t_0$ is negative, 
i.e.~de Sitter expansion {\it has already begun}.  \label{auto3}}
\end{figure}

Finally, the epoch denoted by "III" is the epoch of future exponential expansion, 
or "future inflation", dominated by residual cosmological constant
$\Lambda$. In this epoch, the scale factor behaves according to the de Sitter law, i.e.
\begin{equation}
\label{des}
R(t) = R_0 \exp (Ht),
\end{equation}
where the effective Hubble constant is given as $H=c \sqrt{\Lambda/3}$. 
The present epoch, denoted by $t_0$ is likely located in proximity to the 
epoch $t_i$, given by the eq.~(\ref{al1}), characterizing the onset
of future inflation (Kardashev 1997), but $t_0 > t_i$. In the Fig.~\ref{auto3}, the 
remaining time $t_i =t_0$ until the future inflation sets in is shown as a function of cosmological constant energy density. As the entire discussion, this applies to the case of global flatness of our domain. Since Hubble constant is still poorly known, limiting values $h=0.8$ and $h=0.5$ are drawn (the latter case has somewhat greater weight according to most modern measurements). We notice that, for the recent measurements of $\Omega_\Lambda \sim 0.7$, the change from the power-law expansion toward the de Sitter state given by (\ref{des}) has occured  $2-4$ Gyr ago 
(dependent on the exact value of $H_0$). 
This is already 
small in comparison with the current age of the universe, as given 
by the eq.~(\ref{cage}), the ages of spiral galaxies,\footnote{which are more likely to harbour intelligent observers than the early Hubble types, due to more advanced chemical evolution processes.} and of Pop I stars. Therefore, it is conceivable that the terrestrial life began, 
interestingly enough, in the power-law, matter-dominated phase of
universal expansion, different from the one we are living in now.
This is not completely unexpected, if we keep in mind the Weak Anthropic 
Principle, as pointed out, by Barrow \& Tipler (1986), Weinberg (1987)
and Efstathiou (1995). 

Prospects for survival of intelligent observers (and life in general)
are bleak in any form of open universe (Davies 1973; Dyson 1979; 
Barrow \& Tipler 1986; Tipler 1986). The entropy of each region 
increase, and
all available sources of energy for information processing (actually negentropy) are inevitably depleted on time-scales varying 
according to the particular open model considered. Of course, 
the exponential expansion is fastest at creating such inhospitable environment, since after it sets in the size of the horizon of each observer is shrinking rapidly (for the exact treatment of horizons, see Ellis \& Rothman 1993). The size of the de Sitter horizon
is given in such universe as 
\begin{equation}
\label{des2}
R_h = \frac{c}{H_0} \Omega_\Lambda^{-0.5} \approx 3.6 \times 10^9 \left(
\frac{\Omega_\Lambda}{0.7} \right)^{-0.5} h^{-1} \; {\rm pc}.
\end{equation}
Structures not gravitationally bound to the Local supercluster are
already streaming across this boundary, which will coincide with the
radius of the visible universe in several Gyr (again, exact value depends
on $H_0$). Some time after that, at the epoch which we can denote by $t_h$, the Local supercluster will remain
effectively the only structure existent for local observers. 
However, the exact sequence of events 
depends on the values of several still poorly known parameters, since
the depletion of galactic sources of energy, like normal stars or 
accreting black holes, depends on the total quantity of gas available within gravitationally bound systems and details of star-formation 
physics (Tipler 1986; Adams \& Laughlin 1997; \'Cirkovi\'c 1999). Therefore, this depletion may well be postponed until some point in the inflationary epoch. 
In other words, we still cannot perceive with 
certainty what will be the position of the "see-saw" between local (galactic) and global (cosmological) processes at the moment information processing ceases in our domain. 
 The point of essential inhospitability of eternally expanding universes has been noted by several authors; for instance, Tipler (1986, 1994) insists that only a special subclass of closed universes satisfies FAH within a domain. The three basic conditions formulated by Tipler (1986) to be satisfied for
indefinite information processing seem to preclude FAH in any sort of open universe, and {\it a fortiori\/} any with non-vanishing vacuum energy density.

\section{Many universes and the strategy of survival}

In the sense discussed above, the quantity $t_h$ represents the {\it terminus ante 
quem\/} for the development and evolution of intelligent 
observers in our universe. It is irrelevant for our present discussion whether
$t_h$ has a sharp value or varies slightly due to the growth of cosmological inhomogeneities so that 
limited regions can remain causally connected at some
time $t_h + \Delta t$ (where $\Delta t$ is 
small in comparison to $t_h$ and to $t_0$). However, in order to establish
most stringent constraints, one may require that chain of occurences
which leads to emergence of intelligent observers (i.e.~formation
of Earth-like planets, formation of primitive lifeforms, etc.) begins
before $t_i$, the transitional epoch to the vacuum-dominated expansion. 
This requirement is most probably satisfied for the case of human
observers (as seen from Fig.\ 1, and data on age of life, Earth, etc.). 

What could the strategy of survival in such circumstances look like?
Since the entropy increase in each domain must necessarily preclude further information processing in such a future inflationary universe, even if all other requirements for sustaining intelligent observers are satisfied, 
the only way of continued
existence of such observers would be to leave the cosmological 
domain where they originated. At first, it sounds like a paradox, but in many contemporary
versions of inflationary cosmology, such "escape from the universe" is theoretically possible. Most significant quantum cosmological hypothesis, from this point of view, is so-called "chaotic inflation", devised by Linde (1983), and subsequently developed by Linde (1988, 1990) and other researchers (e.g.~Vilenkin 1995; Garcia-Bellido, 
Garriga \& Montes 1998); it is also discussed in this context by Barrow \& Tipler (1986).

As Linde (1983) writes in the original paper on chaotic inflation:
\begin{squote}
In any case, in the infinite (open) universe at $t \sim t_p$ there should exist infinitely many domains of the type desired, which give rise to an infinite number of mini-universes in which life may exist.
\end{squote}
Many different (but generally similar) concepts of many-universe cosmologies appeared in the last few years, some based on the results of
quantum cosmology, others purely speculative. The details, however,
are not important in our present context. One constraint that the FAH does impose on this type of scenario is that the 
time $t_i$ that is available in the universe before the onset of its exponential 
expansion phase must be sufficient to allow evolution of intelligent life and the 
technological advance necessary for development and application of 
"inter-universe"
transportation, through traversable wormholes or otherwise (e.g.~Morris, Thorne \& Yurtsever 
1988). We thus have the following inequality:
\begin{equation}
\label{cent}
t_i(\Omega, \Lambda, H_0)  \geq \tau.
\end{equation}
The left-hand side of this inequality is cosmological "marble", dependent only on 
the three listed cosmological "parameters" (hopefully to be determined very soon and 
thus cease to be theoretical parameters) while the right-hand side is "wooden" 
evolutionary biology, sociology, technology and other "soft" disciplines (to 
paraphrase Einstein on his field equations in the General Relativity). If we somehow 
knew it were an exact equality, we should have wonder about the strangeness of such 
a coincidence. As it is, this inequality represents a necessary condition for FAH to be correct.

\section{Discussion}
It seems clear that within a single 
cosmological domain, the presence of a non-zero vacuum energy density makes 
indefinite survival of intelligence impossible. However, there are now several theories 
according to which individual domains are but infinitesimally small regions of the 
{\it multiverse}, over which quantum fields vary in a chaotic manner. Some of 
these theories appear to permit travel from one domain to another. Since such theories 
appear quite viable, at least at today's level of understanding, it is still not possible to 
rule out indefinite survival of intelligent beings in this wider context. Thus, while the 
Final Anthropic Principle in its classical formulation pertaining to a single universe can be considered refuted by empirical discovery of $\Omega_\Lambda > 0$, it can be {\it reformulated\/} to encompass the entire multiverse. 

On the other hand, this reformulation may introduce additional problems, of the sort which has been known for the long time, and which faces all cosmologies containing past temporal infinities.
Namely, as succinctly pointed by Paul Davies in his critiques of Ellis' cosmological model (see Davies 1978),
\begin{squote}
There is also the curious problem of why, if the Universe is infinitely
old and life is concentrated in our particular corner of the cosmos, it is not inhabited by technological communities of unlimited age. 
\end{squote}
This problem has been plaguing all variants of the classical steady-
state theory of Bondi \& Gold (1948) and Hoyle (1948).\footnote{There are galaxies of arbitrary age in this theory, since new ones are perpetually created as the old ones recede away, and the characteristic time scale $\tau_s= (3H_0)^{-1}$ is only an average age of galaxies. At the time of formulation of the steady-state theory, $\tau_s$ was considered small ($\sim 6 \times 10^8$ yrs, due to the gross overestimate of the value of Hubble constant at the time), and the Milky Way has already been an extraordinary old galaxy, which certainly implied that surrounding 
galaxies are far less probable to achieve the same degree of  
chemical and biological evolution (if our case 
is an average one). However, if we take $\tau_s$ to be an order of magnitude higher, in accordance of today's best knowledge, 
the anthropic argument of Davies quoted above gains force. Since the fraction of galaxies that are older than age $t$ is given by $\exp (- t/ \tau_s)$, it follows that there are almost 2\% of all galaxies in any 
large enough comoving volume which are {\it twice\/} the age of the 
Milky Way.} Essentially the 
same problem appears if we reformulate Davies' statement in terms of the multiverse: 
Why has our universe (our inflated bubble in Linde's picture) not been colonized or 
visited by some supercivilization originating in a bubble much older than our own?

In the multiverse reformulation, there is now the problem of why our causally connected domain (or even more general, our inflated bubble in Linde's picture) has not already been inhabited (or visited) by 
supercivilizations originating in any bubble much older than ours. 
The simplest solution is to assume that 
interbubble migration is impossible (which might be supported by independent physical evidence in due course). As we have seen above, however, this involves sacrificing the FAH.
While, undoubtedly, there 
are more subtle ways to retain interbubble travel while not producing recognizable effects of technological origin
("The Great Silence"; see e.g.~Brin 1983 and Hanson 
1998), this puzzle further reduces the probability that the FAH is a correct physical proposition.

One conceivable way out for the proponent of FAP would be if it 
turns out that new bubbles are forming at a sufficiently high rate. Associated with travelling from one bubble to another there 
is presumably some minimum cost, such as the matter and 
energy that have to be used in the construction of a wormhole. 
Also, we can assume that it takes some finite time for the 
construction to be completed. A supercivilization contemplating 
colonizing other bubbles would have to choose some strategy of how 
to expend the resources they command. For example, there might be a tradeoff between the amount of energy expended in the construction 
of a wormhole and how quickly the project can be completed. 
On the other hand, it might be worth spending a 
lot of resources to complete a few wormholes quickly in order 
to then use the resources in the domains thus colonized to 
build yet more wormholes, and so on. Call a strategy optimal if it maximizes the expansion rate $dN_C (t)/ dt$ of the civilization in 
the long run (where $N_C(t)$ denotes the number of domains the 
civilization has colonized by time $t$). Suppose that the 
average growth rate of the number of domains in existence 
is $dN_D(t)/dt$. We can then formulate the following inequality:
\begin{equation}
\label{fin}
\frac{dN_C(t)}{dt} < \frac{dN_D(t)}{dt}.
\end{equation}
If this inequality is satisfied, the fraction of all domains that are inhabited by a 
supercivilization originating from some other domain will tend to be negligible, 
consistent with the observation that our universe does not seem to have been 
colonized by any such a civilization. (If, on the other hand, the inequality were 
violated, the multiverse would become satiated with supercivilizations, in apparent 
contradiction to observation.)

Eq.~(\ref{fin}) imposes an upper bound on $dN_D(t)/dt$. The FAH, as 
we saw in earlier sections, implies that $dN_D(t)/dt > 0$, but only 
future research will tell whether FAH is in fact correct.

\acknowledgements{M. M. \'C. wishes to express his gratitude to Milica Topalovic for inspiration and wholehearted support, to Olga Latinovic for help in finding several important references, and to Srdjan Samurovic for some very useful discussions.}

\section*{References}
\refe Adams, F. C. \& Laughlin, G. 1997, Rev. Mod. Phys., 69, 337 

\refe Balashov, Yu. V. 1991, Am. J. Phys., 59, 1069 

\refe Barrow, J. D. 1988, Q. Jl. R. astr. Soc. 29, 101

\refe Barrow, J. D. \& Tipler, F. J. 1986, The Anthropic Cosmological Principle (Oxford University Press, New York)

\refe Bondi, H. \& Gold, T. 1948, MNRAS, 108, 252

\refe Bostrom, N. 1999, in preparation (preprint at {\tt http://www.anthropic-prin- \\ciple.com/preprints.html})

\refe Brin, G. D. 1983, Q. Jl. R. astr. Soc., 24, 283

\refe Carroll, S. M., Press, W. H., \& Turner, E. L. 1992, ARAA, 30, 499

\refe Carter, B. 1974, in Leslie, J. 1990. (edt.) Physical Cosmology and Philosophy (Macmillan Publishing Company)

\refe \'Cirkovi\'c, M. M. 1999, Serb. Astron. J., 159, 79

\refe Davies, P. C. W. 1973, MNRAS, 161, 1

\refe Davies, P. C. W. 1978, Nature, 273, 336

\refe Dyson, F. 1979, Rev. Mod. Phys., 51, 3

\refe Efstathiou, G. 1995, MNRAS, 274, L73

\refe Ellis, G. F. R. \& Rothman, T. 1993, Am. J. Phys. 61, 883

\refe Feynman, R. P. 1965, The Character of Physical Law (Cox and Wyman Ltd., London)

\refe Garcia-Bellido, J., Garriga, J., \& Montes, X. 1998, Phys. Rev. D,
57, 4669 

\refe Guth, A. H. 1981, Phys. Rev. D, 23, 347

\refe Hanson, R. 1999, in preparation (preprint at {\tt http://hanson.berkeley.edu/ greatfilter.html})

\refe Hoyle, F. 1948, MNRAS, 108, 372

\refe Kardashev, N. S. 1997, AZh, 74, 803 

\refe Klapdor, H. V. \& Grotz, K. 1986, ApJ, 301, L39

\refe Leslie, J. 1989, Universes (Routledge, London)

\refe Linde, A. D. 1983, Phys. Lett. B, 129, 177

\refe Linde, A. D. 1988, Phys. Lett. B, 211, 29

\refe Linde, A. D. 1990, Inflation and Quantum Cosmology (Academic 
Press, San Diego)

\refe Martel, H., Shapiro, P. R., \& Weinberg, S. 1998, ApJ, 492, 29

\refe Morris, M. S., Thorne, K. S., \& Yurtsever, U. 1988, Phys. Rev. Lett., 61, 1446

\refe Peebles, P. J. E. 1993, Principles of Physical Cosmology 
(Princeton University Press, Princeton)

\refe Perlmutter, S., Aldering, G., Della Valle, M. et al. 1998, Nature, 391, 51 

\refe Reiss, A. G., Filippenko, A. V., Challis, P. et al. 1998, AJ, 116, 1009 

\refe Rosen, J. 1991, Found. Phys., 21, 977

\refe Singh, A. 1995, Phys. Rev. D, 52, 6700

\refe Tipler, F. J. 1986, Int. J. Theor. Phys. 25, 617

\refe Tipler, F. J. 1994, The Physics of Immortality (Doubleday, New York)

\refe Vilenkin, A. 1995, Phys. Rev. Lett., 74, 846 

\refe Weinberg, S. 1972, Gravitation and Cosmology (Wiley, New York) 

\refe Weinberg, S. 1987, Phys. Rev. Lett., 59, 2607

\end{document}